\documentclass[aps,prl,amsmath,amssymb,superscriptaddress,footinbib,longbibliography,twocolumn]{revtex4-2}
\usepackage{graphicx}
\usepackage[colorlinks,linkcolor=blue,anchorcolor=blue,citecolor=black,urlcolor=black]{hyperref}
\usepackage{amsmath}

\usepackage{color}
\usepackage{soul}
\usepackage{ulem}

\begin{document}
\linespread{1.2}\selectfont  

\title{Observation of quantum strong Mpemba effect}

\author{Jie Zhang}
\thanks{These authors contributed equally to this work.}
\affiliation{Institute for Quantum Science and Technology, College of Science, National University of Defense Technology, Changsha 410073, China}
\affiliation{Hunan Key Laboratory of  Mechanism and technology of Quantum Information, Changsha 410073, China}
\affiliation{Hefei National Laboratory, Hefei 230088, Anhui, China}

\author{Gang Xia}
\thanks{These authors contributed equally to this work.}
\affiliation{Institute for Quantum Science and Technology, College of Science, National University of Defense Technology, Changsha 410073, China}

\author{Chun-Wang Wu}
\affiliation{Institute for Quantum Science and Technology, College of Science, National University of Defense Technology, Changsha 410073, China}
\affiliation{Hunan Key Laboratory of  Mechanism and technology of Quantum Information, Changsha 410073, China}
\affiliation{Hefei National Laboratory, Hefei 230088, Anhui, China}

\author{Ting Chen}
\affiliation{Institute for Quantum Science and Technology, College of Science, National University of Defense Technology, Changsha 410073, China}
\affiliation{Hunan Key Laboratory of  Mechanism and technology of Quantum Information, Changsha 410073, China}
\affiliation{Hefei National Laboratory, Hefei 230088, Anhui, China}

\author{Qian Zhang}
\affiliation{Key Laboratory of Low-Dimensional Quantum Structures and Quantum Control of Ministry of Education, Hunan Normal University, Changsha 410081, China}

\author{Yi Xie}
\affiliation{Institute for Quantum Science and Technology, College of Science, National University of Defense Technology, Changsha 410073, China}
\affiliation{Hunan Key Laboratory of  Mechanism and technology of Quantum Information, Changsha 410073, China}
\affiliation{Hefei National Laboratory, Hefei 230088, Anhui, China}

\author{Wen-Bo Su}
\affiliation{Institute for Quantum Science and Technology, College of Science, National University of Defense Technology, Changsha 410073, China}

\author{Wei Wu}
\affiliation{Institute for Quantum Science and Technology, College of Science, National University of Defense Technology, Changsha 410073, China}
\affiliation{Hunan Key Laboratory of  Mechanism and technology of Quantum Information, Changsha 410073, China}
\affiliation{Hefei National Laboratory, Hefei 230088, Anhui, China}

\author{Cheng-Wei Qiu}
\affiliation{Department of Electrical and Computer Engineering, National University of Singapore, Singapore, Singapore}

\author{Ping-Xing Chen}
\affiliation{Institute for Quantum Science and Technology, College of Science, National University of Defense Technology, Changsha 410073, China}
\affiliation{Hunan Key Laboratory of  Mechanism and technology of Quantum Information, Changsha 410073, China}
\affiliation{Hefei National Laboratory, Hefei 230088, Anhui, China}

\author{Weibin Li}\email{weibin.li@nottingham.ac.uk}
\affiliation{School of Physics and Astronomy, University of Nottingham, Nottingham NG7 2RD, United Kingdom}
\affiliation{Centre for the Mathematics and Theoretical Physics of Quantum Non-equilibrium Systems, University of Nottingham, Nottingham NG7 2RD, United Kingdom}

\author{Hui Jing}
\email{jinghui73@foxmail.com}
\affiliation{Key Laboratory of Low-Dimensional Quantum Structures and Quantum Control of Ministry of Education, Hunan Normal University, Changsha 410081, China}
\affiliation{College of Science, National University of Defense Technology, Changsha 410073, China}

\author{Yan-Li Zhou}\email{ylzhou@nudt.edu.cn}
\affiliation{Institute for Quantum Science and Technology, College of Science, National University of Defense Technology, Changsha 410073, China}
\affiliation{Hunan Key Laboratory of  Mechanism and technology of Quantum Information, Changsha 410073, China}
\affiliation{Hefei National Laboratory, Hefei 230088, Anhui, China}

\date{\today}
\begin{abstract}
	An ancient and counterintuitive phenomenon known as the Mpemba effect showcases the critical role of initial conditions in relaxation processes. How to realize and utilize this effect for speeding up relaxation is an important but challenging task in purely quantum system till now. Here, we experimentally study the strong Mpemba effect in a single trapped ion system in which an exponentially accelerated relaxation in time is observed by preparing an optimal quantum initial state with no excitation of the slowest decaying mode. Also, we demonstrate that the condition of realizing such effect coincides with the Liouvillian exceptional point, featuring the coalescence of both the eigenvalues and the eigenmodes of the systems. Our work provides an efficient strategy to engineer the dynamics of open quantum system, and suggests a link unexplored yet between the Mpemba effect and the non-Hermitian physics.

\end{abstract}

\maketitle

\section{introduction}
\begin{figure*}[htb]
	\includegraphics[scale=1.]{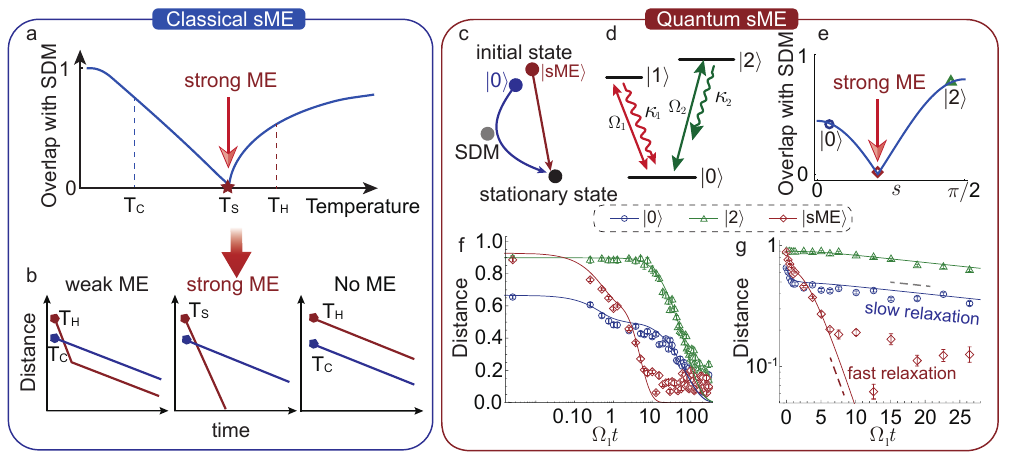}
	\caption{Comparison for classic and quantum Mpemba effect. \textbf{a} The ME can be understood in an intuitive way: the amplitude of the overlap of the initial state with the slowest decaying mode (SDM) depends on the initial temperature in a nonmonotonic way. The sME appears when the overlap with the SDM vanishes. \textbf{b} Weak ME: If an initial high temperature state has a smaller SDM amplitude than that of the lower temperature state, it can reach the thermal equilibrium faster. Strong ME: the system reaches equilibrium at an exponentially faster rate. No ME: the initial high temperature state has a larger overlap with the SDM and thus reaches the equilibrium slower. \textbf{c} By applying a unitary operation, one can realize an initial sME state and approaches the stationary state with a faster rate. \textbf{d} The energy levels for observing the sME (with $\kappa_2\ll\kappa_1$). \textbf{e} The overlap $|c_1|$ of a rotated initial random state with the SDM as a function of the rotation angle $s$. \textbf{f} The distance between the time-relaxed state $\rho(t)$ and the stationary state $\rho_{ss}$ for different initial states: $|0\rangle$ (blue), $|2\rangle$ (green) and $|\mathrm{sME}\rangle$ (red), respectively. The initial sME state starts with a longer distance from $\rho_{ss}$ than the initial state $|0(2)\rangle$ but reaches  $\rho_{ss}$ faster. \textbf{g} The logarithmic scale of the distance evolves with time for different initial states. An exponential speed-up of relaxation is clearly observed for the sME initial state. The experimental parameters for (\textbf{e}-\textbf{g}) are $\Omega_1 =2\pi\times 20 $ kHz, $\Omega_2 = 0.06\Omega_1, \kappa_1 = 2\Omega_1, \kappa_2 = 0.0015\Omega_1$, and the solid lines here are the theoretical predictions based on the experimental parameters.}
	\label{fig:scheme}
\end{figure*}

Relaxations or dissipative evolutions from initial states to a stationary state, widely existing in nature, are vital for fundamental studies of nonequilibrium phenomena and practical control of dynamical devices \cite{Breuer2007,Mohseni2014}. In quantum realm, rapid relaxations are highly desirable for efficient quantum state preparation and qubit engineering \cite{Verstraete2009,Harrington2022,Dann2019,Wu2021c,Khandelwal2021,Kumar2022}. As a possible strategy to achieve this goal, the Mpemba effect (ME) \cite{Mpemba_1969}, well-known in the counterintui-tive example that water can cool faster when initially heated up, has attracted growing interests both in classical  \cite{Lu2017, Klich2019, Gal2020, Kumar2020b, Bechhoefer2021, Teza2023a, Pemartin2024,Ibanez2024} and quantum systems in recent years~\cite{Carollo2021, Nava2019, Kochsiek2022, Manikandan2021,Ares2023,Rylands2023,Ivander2023, Chatterjee2023, Chatterjee2023b,Moroder2024,Boubakour2024,Joshi2024, Shapira2024}. Often, this and related phenomena admits a general explanation \cite{Lu2017,Lasanta2017,Gal2020,Klich2019}: the state of the hotter system overlaps less with the slowest decaying mode (SDM) of the dissipative or cooling dynamics, implying the critical role of initial conditions in relaxations (see Fig. \ref{fig:scheme}(a-b)).

For purely quantum systems at zero temperature, the main challenge is to identify ME-induced rapid relaxations that are not smeared out by quantum superposition~\cite{Carollo2021}. Very recently, Carollo et al. ~\cite{Carollo2021} proposed that strong ME (sME) or exponential speed-up of relaxation can emerge in Markovian open quantum systems by devising an optimal initial state (i.e., sME state) to prohibit excitation of the slowest decaying mode (SDM, see Fig. \ref{fig:scheme}(c)). This prediction of quantum sME, however, has not been experimentally realized till now, hindering its possible applications in e.g. ‘engineered’ relaxation dynamics of the open quantum system \cite{Bechhoefer2021}.

Here, we report the observation of the sME in a truly quantum system, which is a genuine quantum effect and cannot be captured by semi-classical methods. As an essential step towards this target, we construct the sME state via efficient gate operations on a single trapped ion and show that with such  special pure state, featuring zero overlap with the SDM, exponential speeding-up of relaxations can be observed (see Fig. \ref{fig:scheme}(d-g)). Also we find that a critical point can appear in our system, separating the regimes with or without exponential acceleration of relaxations, which coincides well with the Liouvillian exceptional point (LEP)  \cite{Bender2007, Chen2017,Hodaei2017, Miri2019, Minganti2019, Naghiloo2019,Chen2021a,Chen2022}. Furthermore, we observe both eigenvalues and eigenmodes coalesce at the LEP in the experiment by measuring the overlaps with the decaying modes. Our findings indicate a possible link unexplored yet between quantum ME and non-Hermitian physics \cite{Ding2016, Zhou2023, Purkayastha2020, Khandelwal2021, Zhang2022, Chen2021a,Chen2022,Bu2023}, which may stimulate more exciting efforts on e.g., engineering quantum sME with higher-order or topological LEPs \cite{Chen2021a,Chen2022,Bu2023}.

\section{Results}
\subsection{Theory of quantum strong Mpemba effect}
To understand the quantum sME, we consider a Lindblad master equation $\dot{\rho} = \mathcal L \rho(t)$, where $\mathcal L$ is the Liouvillian superoperator \cite{Macieszczak2016a, Carollo2021, Minganti2019}
\begin{eqnarray}
	\mathcal L \rho = -i[H, \rho] + \sum_{\alpha}(J_{\alpha}\rho J_{\alpha}^{\dagger}-\frac{1}{2}\{J_{\alpha}^{\dagger}J_{\alpha}, \rho \}).
\end{eqnarray}
Here $H$ is the Hamiltonian of the system and $J_{\alpha}$ are the quantum jump operators. The density operator $\rho(t)$
can be expanded as the sum of all eigenmodes ($R_i$) of $\mathcal L$
\begin{equation}
	\rho(t) = e^{\mathcal L t} = \rho_{ss} +  \sum_{i=1}^{d^2-1} c_i e^{\lambda_i t} R_i.
\end{equation}
Here $R_i(L_i)$ are the right (left) eigenmatrices of the Liou-villian superoperator $\mathcal L$, with the corresponding eigenvalues $\lambda_0 > \mathrm{Re}[\lambda_1] \ge \mathrm{Re}[\lambda_2] \ge \mathrm{Re}[\lambda_3] \ge ...$, and $\lambda_0=0$. $\lambda_0$ or its eigenmatrix $R_0$ denotes the stationary state $\rho_{ss}$, which is independent of any initial state $\rho_{in}$, while the real parts of other eigenvalues $\lambda_{i\geq 1}$ indicate the relaxation rates of the eigenmodes $R_i$. Coefficients $c_i = \mathrm{Tr}[L_i \rho_{in}]$ give the overlap of $L_i$ with $\rho_{in}$, and $d^2$ denotes the number of the decay modes.

Generally, an initial state can overlap with all decaying modes of Lindblad dynamics, but at long times the relaxation is dominated by the slowest one $R_1$ \cite{Minganti2018,Macieszczak2016a,Teza2023a, Kumar2020b}. The decay rate of the eigenmode $R_1$ sets an exponential timescale of the relaxation $\tau_1 = 1/|\mathrm{Re}[\lambda_1]|$\cite{Znidaric2015,Minganti2018,Macieszczak2016a}, which is normally independent of the initial state. But for the ME case, anomalous fast relaxations can be achieved by designing a special form of the overlap $c_1$ featuring smaller or even zero overlap with the SDM \cite{Klich2019,Kumar2020b,Carollo2021,Kochsiek2022}.

The quantum sME can be realized by designing such an initial state $|\mathrm{sME}\rangle$ satisfying~\cite{Carollo2021},
\begin{eqnarray}
	\mathrm{Tr}[L_1 |\mathrm{sME}\rangle\langle \mathrm{sME}|] = c_1 =0.
\end{eqnarray}
This optimal state $|\mathrm{sME}\rangle$ is prepared by applying a well devised unitary transformation to an initially pure quantum state of the system ~\cite{Carollo2021}. Therefore the sME state is normally a quantum superposition state, which represents a fundamental difference from classic sME, and thus the resulting dynamical behaviour cannot be captured by semi-classical approaches (see Supplementary note 1). Since this sME initial state has zero overlap with the SDM, the relaxation rate of the system is thus $|\mathrm{Re}[\lambda_2]|$, with the timescale $\tau_2 = 1/|\mathrm{Re}[\lambda_2]|$, instead of $\tau_1 = 1/|\mathrm{Re}[\lambda_1]|$. This implies an exponentially faster convergence to the stationary state by a factor $\mathrm{Re}[\delta\lambda_{12}]= \mathrm{Re}[\lambda_1-\lambda_2]$.

\subsection{Experimental approach}

The experimental setup and relevant energy levels for the quantum sME are shown in Fig. 2(a-b). The ground state $|0\rangle = |4^2S_{1/2},m_j=-1/2\rangle$ is resonantly coupled to state $|1\rangle = |3^2D_{5/2},m_j=-5/2\rangle$ and $|2\rangle = |3^2D_{5/2},m_j=3/2\rangle$ by one 729 nm laser beam with two frequency components and the corresponding Rabi frequencies are $\Omega_1$ and $\Omega_2$, respectively. Another laser at 854 nm with right circular polarization induces a tunable decay channel between state $|1 \rangle$ and $|0 \rangle$ with decay rate $\kappa_1$, by coupling $|1\rangle$ to a short-life level $|4^2P_{3/2},m_j=-3/2\rangle$, which will decay quickly back to the state $|0\rangle = |4^2S_{1/2},m_j=-1/2\rangle$ (see Supplementary note 2). The imperfect polarization of 854 nm will cause slow decay from state $|2\rangle$ to $|0 \rangle$ with decay rate $\kappa_2 \ll \kappa_1$ and also the leakage of population to the other Zeeman ground state. Fortunately, the leakage problem can be fixed by introducing a weak optical pumping beam at 397 nm during the data acquisition process. Then for the effective three-level system, we have the Hamiltonian $H=\sum_{j = 1,2}\Omega_j/2(|0\rangle \langle j|+|j\rangle \langle 0|)$ and two jump operators $J_j = \sqrt{\kappa_j}|0\rangle \langle j|$  ($j = 1, 2$).

\begin{figure*}[htb]
	\includegraphics[scale=1.]{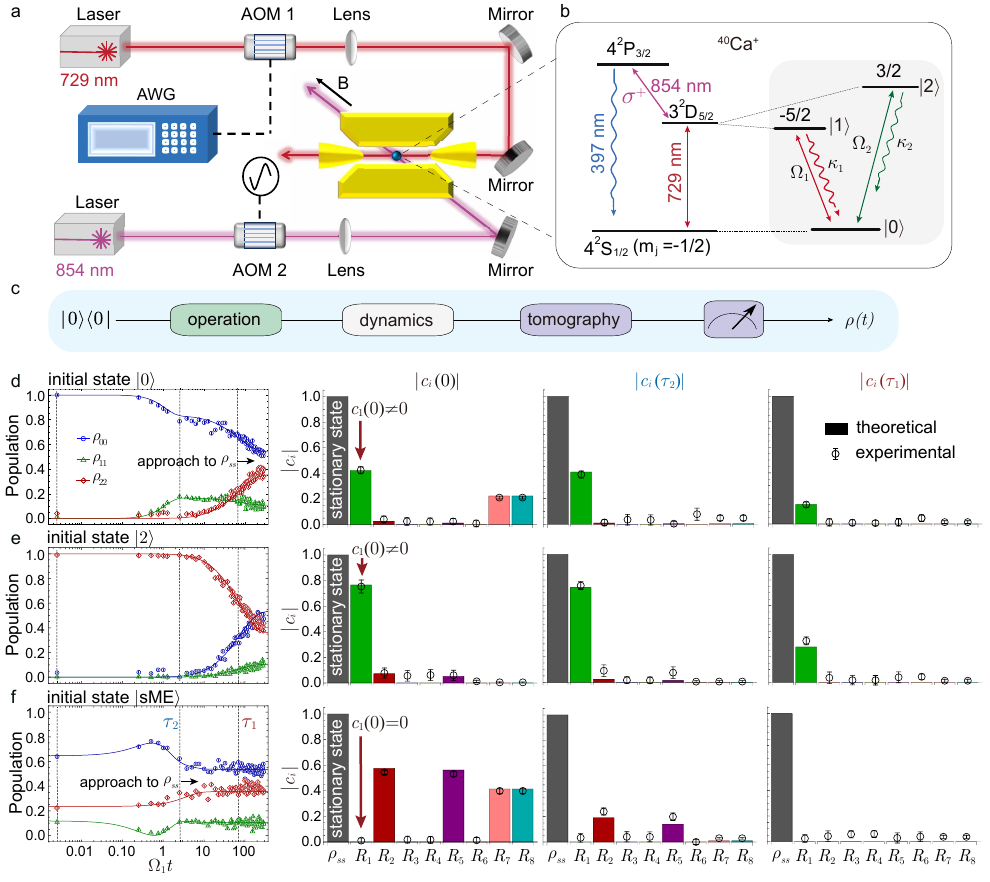}
	\caption{Experimental setup and relaxation dynamics for different initial states. \textbf{a} Experimental setup of quantum sME. The coherent driving between the S state and D state is realized by using a 729 nm laser beam with linewidth about 100 Hz. Two transitions required (as shown in \textbf{b}) are simultaneously driven by injecting two RF frequencies to an acoustic-optic modulator (AOM1) via an arbitrary waveform generator (AWG). The  decay rate of the D state is controlled by  the power of 854 nm laser. \textbf{b} The relevant energy levels of a single $^{40}$Ca$^{+}$ ion involved in the experiment, with state $|0\rangle, |1\rangle, |2\rangle$ are encoded in the energy levels $|4^2S_{1/2},m_j=-1/2\rangle$, $|3^2D_{5/2},m_j=-5/2\rangle$ and $|3^2D_{5/2},m_j=3/2\rangle$, respectively. The energy gap  of state $|1\rangle$ and $ |2\rangle$ can minimize detrimental effect from the polarization impurity of the 854 nm laser beam. \textbf{c} Protocol to generate the sME initial state and tomography of the density matrix $\rho(t)$. After generating the initially sME state by applying two rotation gates on the state $|0\rangle$, the system evolves with the Liouvillian superoperator $\mathcal L$ of interest. After a time $t$, projective measurements are performed for the tomography of the state $\rho(t)$. \textbf{d}-\textbf{f} Left: The dynamics of the state population $\rho_{ii}=\mathrm{Tr}[|i\rangle i\langle|\rho(t)] (i = 0,1,2)$ for initial states $|0\rangle$ (\textbf{d}), $|2\rangle$ (\textbf{e}) and $|\textrm{sME}\rangle$ (\textbf{f}) on a logarithmic timescale. Right: The evolving overlap $|c_{i}(t)|= |\mathrm{Tr}[L_i \rho(t)]|$ for three time stamps at $0, \tau_2, \tau_1$. The solid lines indicate the theoretical results based on the experimental parameters same with Fig. \ref{fig:scheme}(\textbf{e}-\textbf{g}) and the error bars are generated by using Monte Carlo simulation.}\label{fig:dynamics}
\end{figure*}

To observe quantum sME, we initialize the ion in the ground state $|0\rangle$ and subsequently rotate it by applying a unitary operation $U$ \cite{Carollo2021} (see Fig. \ref{fig:dynamics}(c)). Although this transformation can be exactly constructed through the decomposition method \cite{Nielsen2011}, it is still challenging to realize it with high fidelity  in the experiment since at least six gate operations are required for our case.  To overcome this obstacle, here we optimize the operations to two qubit-rotations (see Methods for technical details). The rotated state then relaxes with the desired Liouvillian superoperator $\mathcal L$. The final step is quantum state tomography, i.e., performing the projective measurements on the time-relaxed state $\rho(t)$ as it relaxes from the initial state $\rho_{in}$ to the final stationary $\rho_{ss}$. We remark that both the dynamical process and the state readout need to be carefully regulated, accompanying with the optimization procedure (see Methods).

\subsection{Quantum strong ME}

With the aid of tomography of $\rho(t)$, we can characterize the relaxation process by using the Hilbert-Schmidt distance \cite{Carollo2021}
\begin{eqnarray}
	\mathcal D(\rho(t),\rho_{ss}) = ||\rho(t) - \rho_{ss}||,
\end{eqnarray}
with the notation $||A|| = \mathrm{Tr}[\sqrt{A^{\dagger}A}]$. As presented in Fig. \ref{fig:scheme}(f), the sME initial state generated by our method is farther from the stationary state than normal initial states $|0(2)\rangle$. This is due to the fact that while $U$ removes the SDM excitation, it also modifies the excitation of the remaining ones \cite{Carollo2021}. Nonetheless, the approach to stationarity for sME initial state is still faster since the SDM ($R_1$) is cut off by $U$ \cite{Carollo2021}. Fig. \ref{fig:scheme}(g) gives the Logarithmic scale of distance $\mathcal D(\rho(t),\rho_{ss})$. Compared with the normal initial states, an exponential speeding-up of the relaxation is reached for the sME initial state, a clear evidence of quantum sME. It can also be observed by other distance measures, for instance, the trace distance \cite{Nielsen2011} or Bures distance \cite{VanVu2021c,Garcia-Pintos2022} (see Supplementary note 1).

Fig. \ref{fig:dynamics}(d-f) illustrates the relaxation dynamics of initial states $|0\rangle$, $|2\rangle$,  and $|\textrm{sME}\rangle$, respectively, under  same experimental parameters. We find that they all reach the same steady state after a sufficiently long time, but the relaxation time of the sME initial state is significantly shorter. For the normal initial states $|0(2)\rangle$ (see Fig. \ref{fig:dynamics}(d-e)), when $t\gg \tau_2$ the system relaxes into a state in the metastable manifold till $\tau_1 = 1/|\mathrm{Re}[\lambda_1]|$ \cite{Macieszczak2016a}. Differently, as seen in Fig. \ref{fig:dynamics}(f), the application of $U$, cutting of the excitation of the SDM, leads to striking faster approach to the steady state with the time scale $\tau_2\ll \tau_1$. In order to visualise their overlaps on each decaying mode, we give all coefficients $c_i(t)$ of $\rho(t)$ decomposition into all the decaying modes at time $t=0$, $t=\tau_2$, and $t=\tau_1$, respectively. As depicted in Fig. \ref{fig:dynamics} (d-e), when time $t=0$, a generic initial state will normally overlap with the slowest one, i.e., $c_1(0) \neq 0$. When time $t > \tau_2$, the overlap of initial state with the short-life excitation modes ($R_{i>1}$) becomes very small, while the SDM ($R_1$) is still relevant and keeps the system away from stationarity for a long time till $t \gg \tau_1$. However, for the sME initial state,  $c_1=0$, therefore,  the relaxation time scale is reduced to $\tau_2$ with a faster decay rate $|\mathrm{Re}[\lambda_2]|$ (see Fig.\ref{fig:dynamics} (f)).  Note that the coefficient on $R_0$ (i.e., the stationary state) has the form $c_0\left( t \right) =\mathrm{Tr}[L_0\rho(t)]=\mathrm{Tr}[\rho(t)]=1$ and consistently remains at 1 for all the time, while the coefficients on other decay modes $c_i\left( t \right) =c_ie^{\lambda _it}=\mathrm{Tr[}L_i\rho _{in}]e^{\lambda _it}\,\,\left( i=1,2... \right)$ decrease with time for the reason $\mathrm{Re}[\lambda_{i}]<0$.

\subsection{From strong ME to weak ME}

\begin{figure*}[htb]
	\includegraphics[scale=1.]{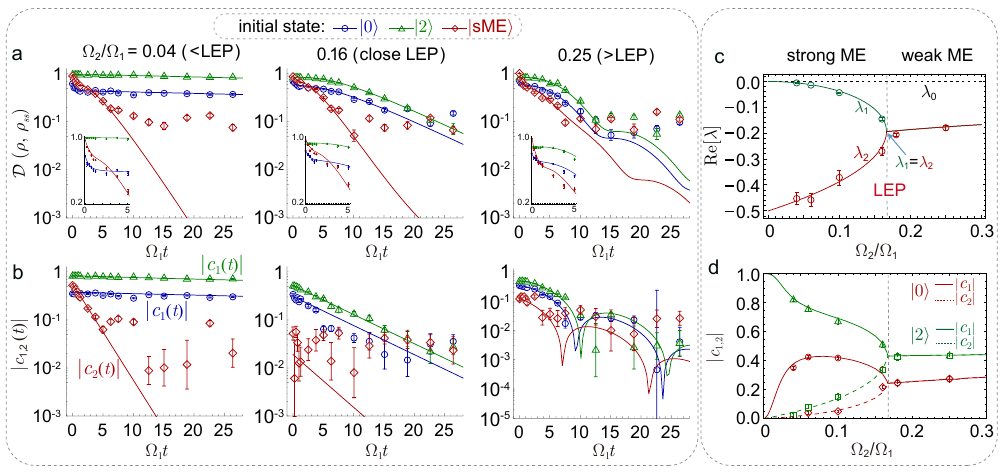}
	\caption{From strong ME to weak ME. \textbf{a} Logarithmic scale of distance $\mathcal{D}(\rho,\rho_{ss})$ evolves with time for different initial state: $|0\rangle$ (blue), $|2\rangle$ (green) and $|\mathrm{sME}\rangle$ (red), respectively. \textbf{b} The corresponding coefficients $c_1 (t)$ (of initial state $|0(2)\rangle$) and $c_2(t)$ (of initial state $|\mathrm{sME}\rangle$). The sME that exits exponential acceleration is observed for $\Omega_2/\Omega_1 = 0.04, \ 0.16\ (< \mathrm{LEP})$. When $\Omega_2/\Omega_1 =0.25\ (> \mathrm{LEP})$, $|\mathrm{sME}\rangle$ has the same decay rate with the normal initial states $|0(2)\rangle$, meaning the strong ME disappears but weak ME is allowed. \textbf{c} Real parts of the eigenvalues of the Liouvillian operator as a function of $\Omega_2/\Omega_1$. \textbf{d} Overlaps $|c_{1}|$ (solid) and $|c_{2}|$ (dashed) for different initial states $|0\rangle$ (blue) and $|2\rangle$ (green), respectively. The parameters are $\Omega_1 =2\pi\times 20 $ kHz, $\kappa_1 = 2\Omega_1, \kappa_2 = 0.0015\Omega_1$. All error bars in this figure are calculated by using Monte Carlo simulation. }
	\label{fig:distance}
\end{figure*}

Can this exponential acceleration always happen for the sME initial state? To check this, we measure the distance $\mathcal D(\rho(t),\rho_{ss})$ and the overlaps $c_{1,2}(t)$ for different parameters of the system. As shown in Fig. \ref{fig:distance}(a-b), for the normal initial state $|0\rangle(|2\rangle)$, the distance has similar dynamical behavior with the overlap $c_1(t)$ at the final stage of the relaxation, i.e., $\rho(t)-\rho_{ss} \simeq c_1 e^{\lambda_1 t}R_1$. While for sME intial state, it becomes to $\rho(t)-\rho_{ss} \simeq c_2 e^{\lambda_2 t}R_2$. When $\Omega_2/\Omega_1 \ll 1$, the accelerated relaxation achieved for sME initial state is very significant, since in this regime $|\mathrm{Re}[\lambda_1]|\ll |\mathrm{Re}[\lambda_2]|$, as shown in Fig. \ref{fig:distance}(c). While as $\Omega_2$ increases, this exponential gain $|\mathrm{Re}[\delta\lambda_{12}]|$ decreases and even disappears when $\Omega_2$ passes across the LEP, where $\lambda_1 = \lambda_2$ and $R_1 = R_2$. The reason is that when $\Omega_2/\Omega_1 > \mathrm{LEP}$, the eigenvalue of the SDM of the $\mathcal{L}$ forms a complex conjugate pair, i.e., $\lambda_1 = \lambda_2^*$ (see Fig. \ref{fig:distance}(c)), which means that the decaying modes $R_1$ and $R_2$ have the same decay rate. Meanwhile, the imaginary parts of eigenvalues $\lambda_{1,2}$ result in the oscillating during the relaxation process \cite{Khandelwal2021,Zhang2022} (Supplementary note 1). Even though, comparing with normal initial states which have two SDMs, the sME initial state here just has one so that it may still be faster to the stationary state. Different with sME featuring a faster decay rate, this acceleration here corresponds to the weak ME which is associated with a smaller overlap with the SDM \cite{Klich2019}. As a consequence, LEP is the boundary between the quantum strong ME and weak ME, showing that an existence of the quantum strong ME is limited to the regime in which $\lambda_1$ is real. It is helpful for deepening the understanding of the relaxation speed limit from both LEPs and quantum ME perspectives.

Based on the measurements of the overlaps, we further demonstrate both eigenvalues (Fig. \ref{fig:distance}(c)) and eigenmodes (Fig. \ref{fig:distance}(d)) coalesce at LEP in the experiment. Whereas the impact of the decaying modes is often limited to transient dynamics, it presents a practical challenge for experimental observation of more than one eigenvalue of $\mathcal L$. Normally, LEP happens at the point that an eigenvalue changes from real to complex \cite{Chen2021a, Chen2022}. However, this change just can show where LEP happens, but does not demonstrate the coalesce of the LEP, because the coalesce of LEPs typically occur in two or more eigenvalues. This challenge could be solved by fitting the measurement of $c_i(t)=c_i e^{\lambda_i t}$ using a single exponential function of time. Based on this, theoretically, one can get the full spectrum of $\mathcal L$. For our system here, considering the overlap $c_2$ is very small for initial state $|0(2)\rangle$), we choose to get $\lambda_2$ by measuring $c_2(t)$ of initial sME state and $\lambda_1$ by measuring $c_1(t)$ of initial state $|0\rangle$. The results, as depicted in Fig. \ref{fig:distance}(c), match well with the theoretical results and show that LEP is signaled by the degenerated of $\lambda_{1,2}.$

As shown in Fig. \ref{fig:distance}(d), the coalesce of the eigenmodes $R_1(L_1)$ and $R_2(L_2)$ at LEP is indicated by $c_1 = c_2$ for a fixed initial state, because at LEP $\mathrm{Tr}[L_1\rho_{in}]=\mathrm{Tr}[L_2\rho_{in}]$. This means the coalesce the coefficients is a detectable way to show the eigenmodes coalesce of LEP. Note that, as we observed in Fig. \ref{fig:distance}(c-d), the bifurcation singularity of the spectrum and the overlaps $c_{1,2}$ is not detectable in the steady state but rather in the relaxation process \cite{Minganti2019}. In fact, for the dynamics at long times, $\rho(t)-\rho_{ss} \simeq c_1 e^{\lambda_1 t}R_1$, thus LEP corresponds the final direction change of the dynamics from $R_1$ to $R_1+R_1^{\dagger}$. This direction change occurring at LEP is the further evidence of a direct link with anomalous phenomena arising in the relaxation process such as sME and the dynamical phase transition \cite{Teza2023a,Orgad2024}.

\section{Discussion}

In summary, we have observed the quantum sME in a single trapped-ion system by preparing an optimal pure state that has zero overlap with the SDM. For the quantum sME initial state, we propose an efficient quantum control technique for state preparation, dynamical engineering and state tomography. In addition, we reveal that quantum sME only happens within a certain parameter range, which is determined by LEP. Together with well-developed techniques of engineering quantum states, our work not only provides a powerful tool for exploring and utilizing the quantum sME as examples of engineered relaxation dynamics \cite{Bechhoefer2021} but also bridges the LEP and quantum ME, two previously independent effects. Besides, the experi-mentally accessible methods discussed in the present manuscript, such as the the efficient unitary transformation and the overlaps measurement, will be valuable tools to assess the quality of the state preparation and the control of the system relaxation. 

Note added: While submitting our manuscript, we became aware of another two experimental studies of the Mpemba effect performed with trapped ion systems appeared on arXiv, and currently they have been published on Phys. Rev. Lett. \cite{Joshi2024, Shapira2024}.

\section{Methods}

\begin{figure*}[ht]
	\includegraphics[scale=0.85]{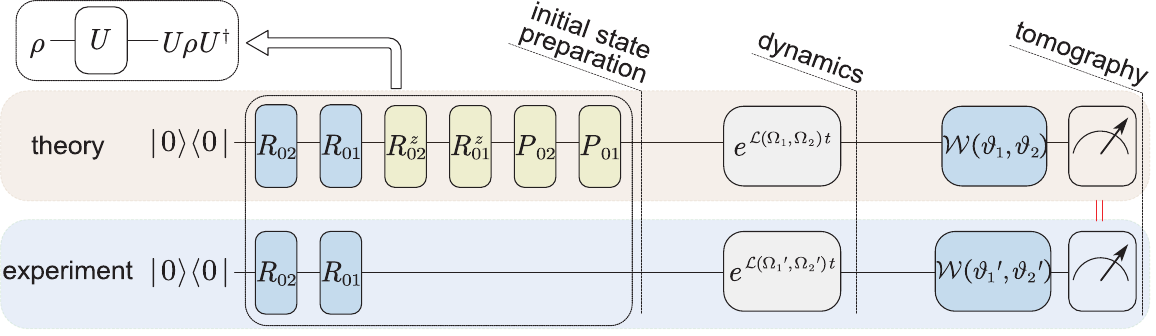}
	\caption{Protocol to prepare arbitrary pure qutrit (three-level) states and to get the density matrix $\rho(t)$ in the experiment. First, we construct the quantum state by engineering an ’virtual’ but efficient unitary transformation $U=R_{02}R_{01}$. Then in the dynamical process, the parameters of the lasers $\Omega_{1,2}$ are adjusted accordingly to $\Omega_1 e^{i\phi}$ and $\Omega_2 e^{i\phi'}$. Finally, the state tomography operations are changed accordingly to $\mathcal W(\vartheta_1',\vartheta_2')$.}
	\label{fig4}
\end{figure*}

We explain here how to devise the unitary transformation $U$ to generate the quantum sME initial (pure) state $\rho_{in}^{\mathrm{sME}} =|\mathrm{sME}\rangle\langle \mathrm{sME}|$. We firstly use the method performed in \cite{Carollo2021} to get $|\mathrm{sME}\rangle$, then we explicitly construct feasible operation $U$ based on our experimental setting (see Fig. \ref{fig4}).

As discussed in \cite{Carollo2021}, the formula of this unitary can be expressed as
\begin{equation}
	U = \exp{[-is(|\phi_1\rangle\langle \phi_2| + |\phi_2\rangle\langle \phi_1|)]|\phi_1\rangle\langle 0|},
\end{equation}
where $|\phi_{1(2)}\rangle$ are the eigenstates of the left-hand eigenmatrix $L_1$ with the corresponding negative (positive) eigenvalues $\alpha_{1(2)}$, $s=\arctan (\sqrt{|\alpha_1/\alpha_2|})$. This qutrit (three-level) $U$ operations can be decomposed into several qubit (two-level) rotations \cite{Nielsen2011}. Compared with qubit operations, a notable challenge in qutrit operations is that the elementary qubit operations lose their ‘global-phase’ gauge freedom because any phase shift is measured relative to the spectator level \cite{Ringbauer2022}.

Now we discuss how to overcome this technical challenge by applying a more efficient and practical way of decomposition, which requires only 2 two-level rotation operations in the experiment (see Fig. \ref{fig4}). Suppose an arbitrary qutrit pure state $|\psi\rangle= U | 0\rangle = (a, b, c)^T$ and $|a|^2+|b|^2+|c|^2 = 1$, then $U$ can be decomposed into a product of 2 two-level unitary matrices $U = AB$ \cite{Nielsen2011}, where
\begin{align}
	A&=\left( \begin{array}{ccc} \frac{a}{\sqrt{|a|^2+|b|^2}} & \frac{b^*}{\sqrt{|a|^2+|b|^2}} & 0 \\ \frac{b}{\sqrt{|a|^2+|b|^2}} & -\frac{a^*}{\sqrt{|a|^2+|b|^2}} & 0 \\ 0 & 0 & 1 \end{array}\right), \nonumber \\ 
	B&=\left( \begin{array}{ccc} a' & 0 & c^* \\ 0 & 1 & 0 \\
		c & 0 & -a'^*  \end{array}\right),
\end{align}
with $a' = \sqrt{|a|^2+|b|^2}$. The unitary $A$ and $B$ can be written in the form
\begin{eqnarray}
	A&=& P_{01}(\alpha) Z_{01}(\beta) R_{01} (\gamma,\pi/2) Z_{01}(\delta)\nonumber\\
	&=& P_{01}(\alpha) R_{01}^z(\beta +\delta ) R_{01} (\gamma ,\pi/2-2\delta ),\nonumber\\
	B&=& P_{02}(\alpha ') Z_{02}(\beta ') R_{02} (\gamma ',\pi/2) Z_{02}(\delta ' )\nonumber\\
	&=&P_{02}(\alpha ') Z_{02}(\beta '+\delta ') R_{02} (\gamma ',\pi/2-2\delta '),
\end{eqnarray}
where
\begin{eqnarray}
	P_{01}(x)=\left(
	\begin{array}{ccc}
		e^{i x} & 0 & 0 \\
		0 & e^{i x} & 0 \\
		0 & 0 & 1 \\
	\end{array}
	\right),
	\ P_{02}(x)=\left(
	\begin{array}{ccc}
		e^{i x} & 0 & 0 \\
		0 & 1 & 0 \\
		0 & 0 & e^{i x} \\
	\end{array}
	\right),\nonumber\\
	\ Z_{01}(x)=\left(
	\begin{array}{ccc}
		e^{-i x} & 0 & 0 \\
		0 & e^{i x} & 0 \\
		0 & 0 & 1 \\
	\end{array}
	\right),
	\ Z_{02}(x)=\left(
	\begin{array}{ccc}
		e^{-i x} & 0 & 0 \\
		0 & 1 & 0 \\
		0 & 0 & e^{i x} \\
	\end{array}
	\right),\nonumber
\end{eqnarray}

\begin{eqnarray}
	R_{01}(x,y)=\left(
	\begin{array}{ccc}
		\cos \left(\frac{x}{2}\right) & -i e^{-i y} \sin \left(\frac{x}{2}\right) & 0 \\
		-i e^{i y} \sin \left(\frac{x}{2}\right) & \cos \left(\frac{x}{2}\right) & 0 \\
		0 & 0 & 1 \\
	\end{array}
	\right),\nonumber\\
	R_{02}(x,y)=\left(
	\begin{array}{ccc}
		\cos \left(\frac{x}{2}\right) & 0 & -i e^{-i y} \sin \left(\frac{x}{2}\right) \\
		0 & 1 & 0 \\
		-i e^{i y} \sin \left(\frac{x}{2}\right) & 0 & \cos \left(\frac{x}{2}\right) \\
	\end{array}
	\right),\nonumber
\end{eqnarray}
and

\begin{align} 
	\gamma & =2 \cos ^{-1}(A_{11}),\nonumber\\
	\alpha &=\frac{1}{2} (\arg (A_{11})+\arg (A_{22}),\nonumber\\
	\beta&=\frac{1}{2} (\arg (A_{21})-\arg (A_{11})),\nonumber\\
	\delta  &=-\alpha  +\arg (A_{22})-\beta,\nonumber\\
	\gamma  '&=2 \cos ^{-1}(B_{11}),\nonumber \\
	\alpha  '&=\frac{1}{2} (\arg (B_{11})+\arg (B_{33})),\nonumber \\
	\beta  '&=\frac{1}{2} (\arg (B_{31})-\arg (B_{11})),\nonumber \\
	\delta  '&=-\alpha  '+\arg(B_{33})-\beta '.\nonumber
\end{align}
Here we have translated the phase gates $P$ and $Z$ backwards which implement appropriate phase-shifts between the qutrit states \cite{Ringbauer2022, McKay2017}. Then the final decomposed unitary operation for $U$ can be written as
\begin{eqnarray}
	U&=&AB\nonumber\\
	&=&P_{01}(\alpha ) Z_{01}(\beta +\delta ) R_{01} (\gamma ,\pi/2-2\delta ) P_{02}(\alpha ')  \nonumber \\
	& &Z_{02}(\beta '+\delta ') R_{02} (\gamma ',\pi/2-2\delta ')   \\
	&=& P_{01}(\alpha ) P_{02}(\alpha ') Z_{01}(\beta +\delta )Z_{02}(\beta '+\delta ')\nonumber \\
	& &R_{01}(\gamma ,\phi )R_{02}(\gamma ',\phi '),\nonumber
\end{eqnarray}
with $\phi =\alpha '-(\beta '+\delta ')+\frac{\pi }{2}-2 \delta , \phi '=\frac{\pi }{2}-2 \delta '$.

We can further simplify the unitary operation $U$ in a more efficient way by translating phase gates $P$ and $Z$ to the operation backwards till the detection due to the fact that the detection does not need phase information \cite{McKay2017}. By calculating the translated phase in software, we just need two rotation operations $R_{01}R_{02}$ to generate arbitrary qutrit pure state experimentally, which can not only greatly simplify operations, but also improve the fidelity of sME state. 

As shown in Fig. \ref{fig4}, our careful analysis shows that the corresponding dynamical operator and state tomography operations need to be modulated accordingly,
\begin{eqnarray}
	\mathcal{L}(\Omega_1, \Omega_2) \rightarrow \mathcal{L}(\Omega_1 e^{i\phi },\Omega_2 e^{i\phi '}).
\end{eqnarray}
The state tomography for the qutrit system here requires 9 measurement basis, which are $|0\rangle$, $|1\rangle$, $|2\rangle$,  $(|0\rangle+|1\rangle)/\sqrt{2}$,$(|0\rangle+i|1\rangle)/\sqrt{2}$, $(|0\rangle+|2\rangle)/\sqrt{2}$, $(|0\rangle+i|2\rangle)/\sqrt{2}$, $(|1\rangle+|2\rangle)/\sqrt{2}$, $(|1\rangle+i|2\rangle)/\sqrt{2}$. However, only S or D state can be detected in the trapped ion system, therefore we need to rotate the basis to state $|0\rangle$ before applying the detection beam since we have encoded state  $|1\rangle$ and $|2\rangle$ in D state manifold.  The rotation operation $\mathcal{W}(\vartheta_1 e^{i\phi_{L1}}, \vartheta_2 e^{i\phi_{L2}})$ in the state tomography can be realized by simply combining the  $\pi$  or $\pi/2$ pulses with translating phase $ \phi_{L1} = \alpha^\prime - 2(\beta + \delta) - (\beta^\prime +\delta^\prime)$ on the transition $|0\rangle \leftrightarrow |1\rangle$ and $ \phi_{L2} = \alpha -2(\beta^\prime + \delta^\prime) - (\beta + \delta)$ on the transition  $|0\rangle \leftrightarrow |2 \rangle$ respectively.  After we get the projective measurement results of these basis, we can reconstruct density matrix by using maximum-likelihood method, and the error bars for the density matrix are obtained by using numerical simulations \cite{PhysRevA.66.012303,hu2023self}.

\section{acknowledgments}
This work was supported by the Natural Science Foundation of Hunan Province of China under grant Nos. 2023JJ30626(Y.L.Z.), 2022RC1194(J.Z.), 2023JJ10052(J.Z.), National Natural Science Foundation of China under grant Nos. 12004430(J.Z.), 12174448(C.W.W.), 12074433(P.X.C.), 12174447(W.W.) and 12204543(T.C.). H.J. is supported by the NSFC (grant No.11935006), the Science and Technology Innovation Program of Hunan Province (grant No. 2020RC4047), National Key R\&D Program of China (No. 2024YFE0102400) and Hunan provincial major sci-tech program (No. 2023ZJ1010), and P.X.C. acknowledges the support from Innovation Program for Quantum Science and Technology under grant No. 2021ZD0301605.

\section{Author contributions}
Y.L.Z. conceived and designed the research with helpful discussions with H.J.,W.L. and P.X.C.; J.Z. and G.X. were responsible for the experimental implementation, the data acquisition and its evaluation; J.Z., Y.L.Z. and C.W.W. designed the experiment with the help of T.C., Y.X., W.W. and P.X.C.; Y.L.Z. performed the simulations; Y.L.Z., J.Z. and G.X. analyzed the data; Y.L.Z., J.Z, H.J., W.L. and C.W.Q. wrote the paper with inputs from G.X., T.C., Q.Z. and W.B.S. ; H.J., W.L. and P.X.C. supervised this project. All authors discussed the results and contributed to the manuscript.

\section{COMPETING INTERESTS}
The authors declare no competing interests.


\clearpage

	\newpage
	\onecolumngrid
	\appendix

	
	\counterwithin{figure}{section}
	\section{Supplementary Information:
		Observation of quantum strong Mpemba effect}
	
	\author{Jie Zhang}
	\thanks{These authors contributed equally: Jie Zhang, Gang Xia }
	\affiliation{Institute for Quantum Science and Technology, College of Science, National University of Defense Technology, Changsha 410073, China}
	\affiliation{Hunan Key Laboratory of  Mechanism and technology of Quantum Information, Changsha 410073, China}
	\affiliation{Hefei National Laboratory, Hefei 230088, Anhui, China}
	
	\author{Gang Xia}
	\thanks{These authors contributed equally: Jie Zhang, Gang Xia }
	\affiliation{Institute for Quantum Science and Technology, College of Science, National University of Defense Technology, Changsha 410073, China}
	
	\author{Chun-Wang Wu}
	\affiliation{Institute for Quantum Science and Technology, College of Science, National University of Defense Technology, Changsha 410073, China}
	\affiliation{Hunan Key Laboratory of  Mechanism and technology of Quantum Information, Changsha 410073, China}
	\affiliation{Hefei National Laboratory, Hefei 230088, Anhui, China}
	
	\author{Ting Chen}
	\affiliation{Institute for Quantum Science and Technology, College of Science, National University of Defense Technology, Changsha 410073, China}
	\affiliation{Hunan Key Laboratory of  Mechanism and technology of Quantum Information, Changsha 410073, China}
	\affiliation{Hefei National Laboratory, Hefei 230088, Anhui, China}
	
	\author{Qian Zhang}
	\affiliation{Key Laboratory of Low-Dimensional Quantum Structures and Quantum Control of Ministry of Education, Hunan Normal University, Changsha 410081, China}
	
	\author{Yi Xie}
	\affiliation{Institute for Quantum Science and Technology, College of Science, National University of Defense Technology, Changsha 410073, China}
	\affiliation{Hunan Key Laboratory of  Mechanism and technology of Quantum Information, Changsha 410073, China}
	\affiliation{Hefei National Laboratory, Hefei 230088, Anhui, China}
	
	\author{Wen-Bo Su}
	\affiliation{Institute for Quantum Science and Technology, College of Science, National University of Defense Technology, Changsha 410073, China}
	
	\author{Wei Wu}
	\affiliation{Institute for Quantum Science and Technology, College of Science, National University of Defense Technology, Changsha 410073, China}
	\affiliation{Hunan Key Laboratory of  Mechanism and technology of Quantum Information, Changsha 410073, China}
	\affiliation{Hefei National Laboratory, Hefei 230088, Anhui, China}
	
	\author{Cheng-Wei Qiu}
	\affiliation{Department of Electrical and Computer Engineering, National University of Singapore, Singapore, Singapore}
	
	\author{Ping-Xing Chen}
	\affiliation{Institute for Quantum Science and Technology, College of Science, National University of Defense Technology, Changsha 410073, China}
	\affiliation{Hunan Key Laboratory of  Mechanism and technology of Quantum Information, Changsha 410073, China}
	\affiliation{Hefei National Laboratory, Hefei 230088, Anhui, China}
	
	\author{Weibin Li}\email{weibin.li@nottingham.ac.uk}
	\affiliation{School of Physics and Astronomy, University of Nottingham, Nottingham NG7 2RD, United Kingdom}
	\affiliation{Centre for the Mathematics and Theoretical Physics of Quantum Non-equilibrium Systems, University of Nottingham, Nottingham NG7 2RD, United Kingdom}
	
	\author{Hui Jing}
	\email{jinghui73@foxmail.com}
	\affiliation{Key Laboratory of Low-Dimensional Quantum Structures and Quantum Control of Ministry of Education, Hunan Normal University, Changsha 410081, China}
	\affiliation{College of Science, National University of Defense Technology, Changsha 410073, China}
	
	\author{Yan-Li Zhou}\email{ylzhou@nudt.edu.cn}
	\affiliation{Institute for Quantum Science and Technology, College of Science, National University of Defense Technology, Changsha 410073, China}
	\affiliation{Hunan Key Laboratory of  Mechanism and technology of Quantum Information, Changsha 410073, China}
	\affiliation{Hefei National Laboratory, Hefei 230088, Anhui, China}

	\subsection{Supplementary Note 1. Markovian open quantum dynamics, Liouvillian exceptional point, and quantum sME}
	
	We first briefly discuss the fundamental elements of open quantum systems evolving under Markovian dynamics. The evolution of the density matrix $\rho(t)$, is generated by the Lindblad master equation
	\begin{equation}
	\frac{d\rho}{dt} = \mathcal L \rho(t):= -i[H,\rho]+\sum_j (2J_j \rho J_j^{\dagger} - \{J_j^{\dagger} J_j, \rho\}),
	\end{equation}
	where $\mathcal L $ is the Liouvillian superoperator, $H$ is the Hamiltonian governing the unitary part of the dynamics, $J_j$ are the jump operators describing the dissipative process. Since the Liouvillian $\mathcal{L}$ acts linearly on $\rho(t)$, one can obtain the time evolution about the relaxation of any initial state $\rho_{in}$ in terms of its eigenmatrices (eigenmodes) $R_i$ and the corresponding complex eigenvalues $\lambda_i$ via the relation $\mathcal L R_i = \lambda_i R_i$:
	\begin{eqnarray}
	\rho(t) = e^{\mathcal L t} \rho_{in} = \rho_{ss} + \sum_{i\geqslant 1}^{d^2-1} c_i e^{\lambda_i t} R_i.
	\end{eqnarray}
	Here $c_i = \mathrm{Tr}[L_i \rho_{in}]$ are coefficients of the initial state decomposed into the eigenmatrices of $\mathcal L^{\dagger}$ with $\mathcal L^{\dagger} L_i = \lambda_i^* L_i$, $R_i$ and $L_i$ are referred as right and left eigenmatrices, respectively, and can be normalized by  Tr$[L_i R_j] = \delta_{ij}$. The stationary state of the system under consideration is then given by the density matrix $\rho_{ss}$ such that $\mathcal L \rho_{ss} = 0$, i.e., $\rho_{ss}=R_0$, which corresponds to the zero eigenvalue $\lambda_0 =0$ and is independent of the initial state. The trace preservation of the dynamics implies that
	$\mathrm{Tr}[\rho(t)]= \mathrm{Tr}[\rho_{ss}] =1 = \mathrm{Tr}[L_0R_0]$, and thus $L_0$ is the identity ($L_0 = I$). It also implies that $\mathrm{Tr}[R_{i\geqslant 1}] =0$, which means other right eigenmatrices do not correspond to quantum states.
	
	If the eigenvalues are ordered by decreasing their real parts, it is known that the negative real parts of the eigenvalues, $\mathrm{Re}[\lambda_{i>0}] < 0$, determine the relaxation rates of the system towards the stationary state, and the corresponding eigenmatrices $R_{i>0}$ are called decaying modes \cite{Znidaric2015}. While the imaginary parts describe the possible oscillatory processes. Note that, due to the Hermiticity preserving of $\mathcal L$, if $\lambda_i$ is complex, $\lambda_i^*$ must also be an eigenvalue of $\mathcal L$ \cite{Minganti2019,Minganti2018,Carollo2021}. As a consequence, the eigenvalues are symmetrically distributed with respect to the real axis. For conjugate pairs, $\lambda_{i,1} = \lambda_{i,2}^*$, the corresponding eigenmodes can be constructed to be Hermitian by considering the combinations $ (R_{i,1}+R_{i,2})$ and $i (R_{i,1}-R_{i,2})$, then their corresponding eigenvalues are real.
	
	A generic initial state will overlap with all decaying modes of Lindblad dynamics, and thus, in particular, also with the slowest one $R_1$, which plays a fundamental role in the system dynamics \cite{Znidaric2015}. Then the Liouvillian gap,  defined by $g = |\mathrm{Re}[\lambda_1]|$, is also called the relaxation rate and thus is an important quantity determining the timescale of the final relaxation to the stationary state $\tau_1 = 1/|\mathrm{Re}[\lambda_1]|$ \cite{Macieszczak2016a,Minganti2018}. This means that the system dynamics at long times has the form
	\begin{eqnarray}
	\rho(t) - \rho_{ss} \simeq c_1 e^{\lambda_1 t} R_1,
	\end{eqnarray}
	and possesses the slowest decay rate $|\mathrm{Re}[\lambda_1]|$. As such, for long times, the system will relax to $\rho_{ss}$ with a purely exponential decay when $\lambda_1$ is real.
	
	For a quantum sME initial state $|\mathrm{sME}\rangle$, which has zero overlap with the slowest decaying mode (SMD) $R_1$, then we will get $c_1 = \mathrm{Tr}[L_1 (|\mathrm{sME}\rangle\langle \mathrm{sME}|)]=0$. Note that this optimal state $|\mathrm{sME}\rangle$ is prepared by applying a well devised unitary transformation to an initially pure quantum state of the system ~\cite{Carollo2021}. Therefore the sME state is normally a quantum superposition state, which is fundamentally different from classic ME. Besides, the relaxation process of our system contains the coherent evolution part, so that even the final steady state has quantum coherence. This can be illustrated by the state tomography of the sME initial state and the steady state, see Supplementary Fig. \ref{fig_s1}. The non-zero value of non-diagonal elements indicates the quantum coherence here. Finally, for the sME initial state, the following dynamical behaviour cannot be captured by semi-classical approach. To show this, we compare the dynamics of the populations by using the rate equations (dashed lines) and the Lindblad master equation (solid lines) and the results are shown in Supplementary Fig. \ref{fig_s2}. Obviously, the dynamics for the sME initial state by using the rate equations (dashed lines) does not cause exponential acceleration. This proves that conventional rate equation approaches cannot capture the quantum Mpemba effect that is observed in our experiment.
	
	\begin{figure}[htb]
	\includegraphics[scale=0.55]{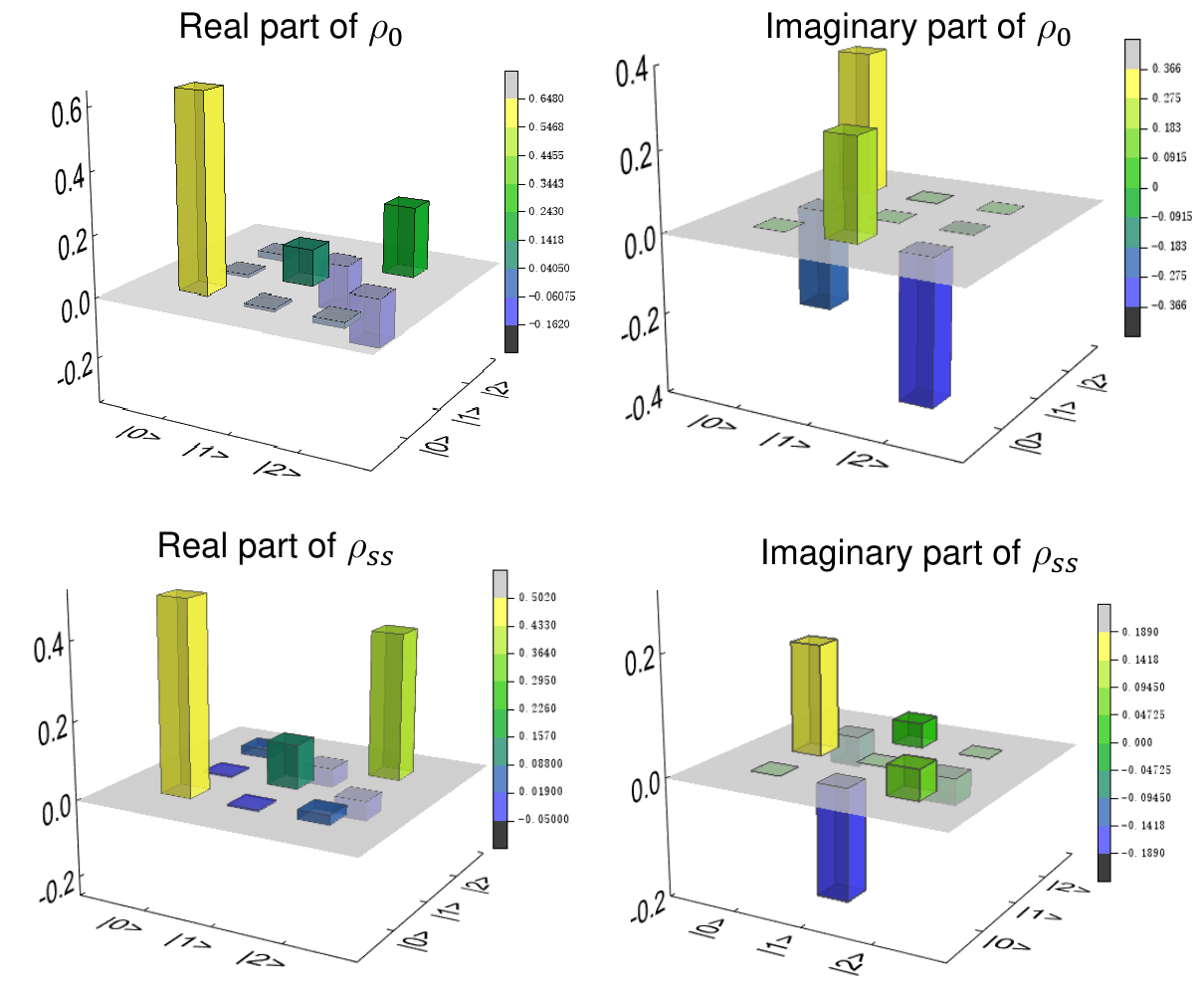}
	\caption{The tomography of the sME initial state $\rho_{0} = |\mathrm{sME}\rangle\langle \mathrm{sME}| $ and the final stationary state $\rho_{ss}$. The parameters are the same with Fig. 2(f).}
	\label{fig_s1}
	\end{figure}
	
	\begin{figure}[htb]
	\includegraphics[scale=1.8]{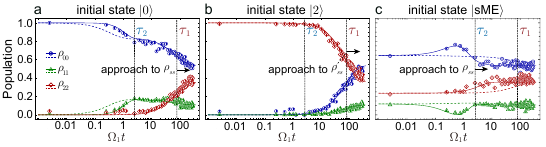}
	\caption{The dynamics of the state populations for initial states $|0\rangle$ (a), $|2\rangle$ (b) and $|\mathrm{sME}\rangle$ (c) on a logarithmic timescale. Dashed lines: the rate equations, solid lines: the Lindblad master equation. The parameters are the same with Fig. 2(d-f). The error bars are generated by using Monte Carlo simulation.}
	\label{fig_s2}
	\end{figure}
	
	Whereas the sME initial state no longer excites the SDM, it generally relaxes with a faster decay rate $|\mathrm{Re}[\lambda_2]|$. If $\lambda_{1,2} \in \mathbb{R}$, the quantum sME initial state will generate an exponential speeding-up of the convergence to stationarity by a factor $\delta \lambda_{12} = \lambda_1 -\lambda_2 > 0$. However, it is impossible to have an
	exponential acceleration if the parameter of the system passes beyond the Liouvillian exceptional point (LEP). The reason is that $\lambda_{1,2}$ becomes complex valued when $\Omega_2/\Omega_1 > \mathrm{LEP}$ and hence $\delta \lambda_{12} = \mathrm{Re}[\lambda_1 - \lambda_2] = 0$. The key point is that if $\lambda_{1} = \lambda_2^*$, then $R_1 = R_2^{\dagger} $ is non-Herminan, namely $R_1 \neq R_1^{\dagger}$ so that $c_1 = c_2^* \neq c_2$.
	
	It also can be explained from long time evolution of $\rho(t)$, where the two lowest decaying mode of the Liouvillian superoperator appear in conjugate paris, i.e.,$\lambda_1 = \lambda_2^*$. The difference between $\rho(t)$ and stationary state $\rho_{ss}$ can be expressed as
	\begin{eqnarray}
	\rho(t) - \rho_{ss} &\simeq & c_1 e^{\lambda_1 t} R_1 +  c_1^* e^{\lambda_1^* t} R_1^{\dagger} \nonumber\\
	&= & |c_1| e^{\mathrm{Re}[\lambda_1] t}( e^{i(\omega_1 t + \delta_1)} R_1 +e^{-i(\omega_1 t + \delta_1)} R_1^{\dagger}) \nonumber\\
	&= & |c_1| e^{\mathrm{Re}[\lambda_1] t}( \cos(\omega_1 t + \delta_1) R_1' +\sin(\omega_1 t + \delta_1) R_2') \nonumber\\
	&= &  c_1(t) R_1' + c_2(t) R_2',
	\end{eqnarray}
	where $\omega_1=\mathrm{Im}[\lambda_1], \delta_1 =\mathrm{Arg}[c_1], c_1(t) = |c_1|e^{\mathrm{Re}[\lambda_1] t}\cos(\omega t +\delta_1), c_2(t) = |c_1|e^{\mathrm{Re}[\lambda_1] t}\sin(\omega t +\delta_1)$, and Hermitian eigenmatrices $R_1' = (R_1 + R_1^{\dagger})$, $R_2' = i(R_1 - R_1^{\dagger})$. Similar to the case of real value of $\lambda_1$, the timescale of relaxation here is still  determined by $|\mathrm{Re}[\lambda_1]|$, but the dynamics which constitutes rotations will be determined by $R_1$ and $R_2$ simultaneously. Thus it can be interpreted as a phase transition in the relaxation dynamics \cite{Teza2023a}. Note that this singularity can not be detected in the steady state but rather in the relaxation dynamics.
	Considering $R_1'$ is Hermitian, we still can generate a pure initial state $|\mathrm{sME}\rangle$ that has zero overlap in
	$R_1'$, i.e., $c_1(0) = \mathrm{Tr}[L_1'(|\mathrm{sME}\rangle\langle \mathrm{sME}|)] = 0 = |c_1| \cos(\delta_1)$, but another SDM $R_2'$, with the same decay rate $|\mathrm{Re}[\lambda_1]|$, cannot be eliminated at the same time because $c_2(0) = |c_1| \sin(\delta_1) \neq 0$. As a consequence, quantum sME will not exist when $\lambda_1$ is complex; the LEP partitioned parameter space shows that the existence of the quantum sME is limited to the region in which $\lambda_1$ is real (see Supplementary Fig. \ref{fig_s3}).We also check that similar results are found using Bures distance \cite{VanVu2021c,Garcia-Pintos2022} and the results are shown in Supplementary Fig. \ref{fig_s4}.
	
	\begin{figure}[htb]
	\includegraphics[scale=0.55]{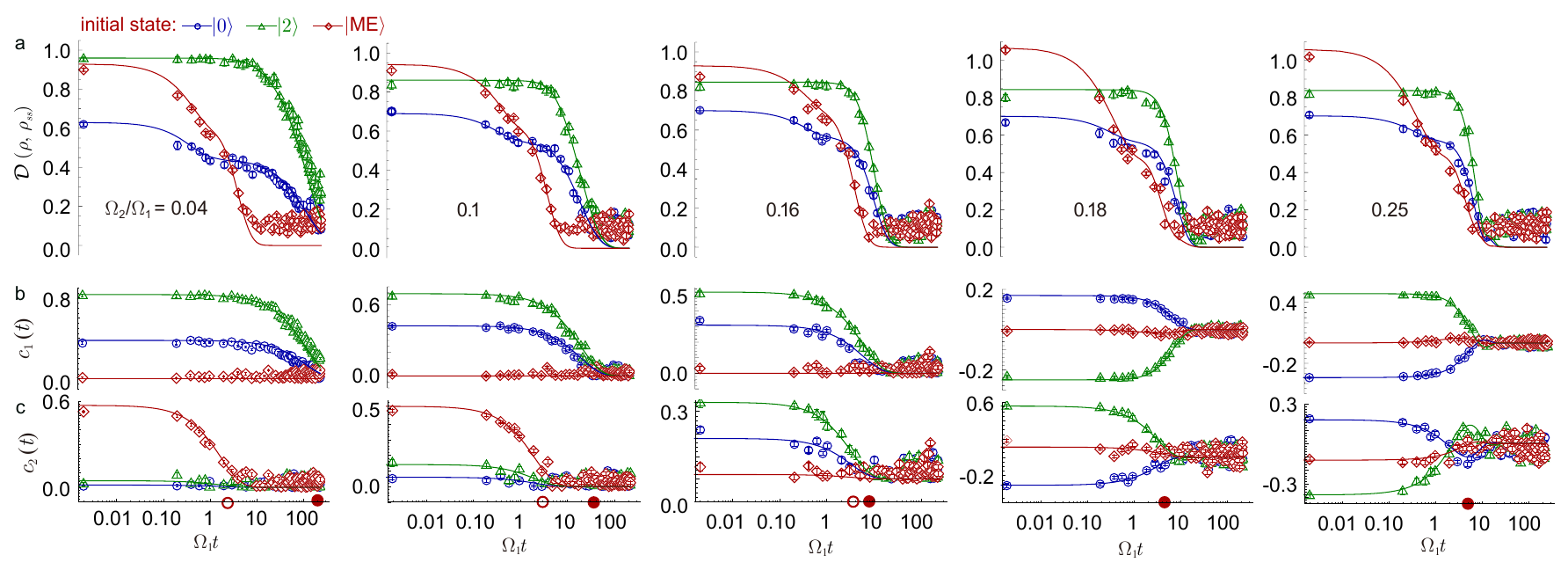}
	\caption{(a) Distance for $\Omega_2/\Omega_1 = 0.04,\ 0.1,\ 0.16,\ 0.18, \ 0.25$, with initial state $|0\rangle$ (green), $|2\rangle$ (blue) and $|\mathrm{ME}\rangle$ (red), respectively. (b-c) The corresponding coefficients $c_1 (t)$ (b) and $c_2(t)$. (c) The sME that exits exponential acceleration is observed for $\Omega_2/\Omega_1 <\mathrm{LEP}$. Considering that $c_{1,2}(t) \in \mathbb{C}$ when $\Omega_2/\Omega_1 < \mathrm{LEP}$ and $c_{1,2}(t) \in \mathbb{R}$ when $\Omega_2/\Omega_1 > \mathrm{LEP}$, for $\Omega_2/\Omega_1 = 0.04,\ 0.1, \ 0.16\ (< \mathrm{LEP})$ what we plot is $|c_{1,2}(t)|$ and for $\Omega_2/\Omega_1 = 0.18,\ 0.25\ (> \mathrm{LEP})$ what we plot is $c_{1,2}(t)$. Here, the open symbols and the filled symbols in the $x$ axis are time scales $\tau_2$ and $\tau_1$, respectively. The error bars are generated by using Monte Carlo simulation.}
	\label{fig_s3}
	\end{figure}
	
	\begin{figure}[htb]
	\includegraphics[scale=1.8]{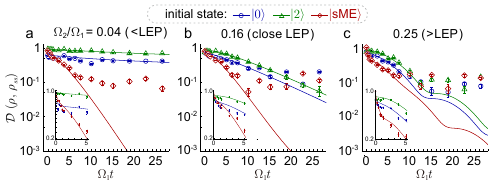}
	\caption{Beres distance for $\Omega_2/\Omega_1 = 0.04$ (a),  0.16 (b), 0.25(c), with initial state $|0\rangle$ (green), $|2\rangle$ (blue) and $|\mathrm{ME}\rangle$ (red), respectively. The parameters are the same with Fig. 3(a) and the error bars are generated by using Monte Carlo simulation.}
	\label{fig_s4}
	\end{figure}
	
	LEP can also imply a discontinuity in the final direction of the approach to the stationary state and generates the dynamical phase transition \cite{Orgad2024,Teza2023a} as a function of the system parameter in the relaxation dynamics (see Fig. 3d). The singularity of this kind of transition, which is induced by the discontinuity of the final direction of the relaxation to stationary state, is intrinsically hard to detect in nature \cite{Teza2023a}. Here we show how this kind of transition can be observed in an experimentally feasible quantum system by measuring the overlap $c_1$ . Because the system dynamics at long times has the form $\rho(t) - \rho_{ss}\simeq c_1e^{\lambda_1 t}R_1$, the final direction will change from $R_1$ to $R_1+R_2$ when the parameter passes through LEP. Obviously, this direction change at LEP will lead to a singularity of $c_1 = \mathrm{Tr}[L_1\rho_{in}]$. For a quantum sME initial state, however, at LEP we have $\lambda_1(R_1) = \lambda_2(R_2)$ so that $c_1 = c_2 =0$. It corresponds to \textit{super-strong} ME \cite{Klich2019} and the timescale of the relaxation is given by the $\tau_3 = 1/|\mathrm{Re}[\lambda_3]|$. This means that at LEP the sME initial state implies an exponentially faster convergence to the steady state by a factor $\mathrm{Re}[\lambda_1-\lambda_3]$, and the final stage of the the relaxation is determined by $R_3$. Therefore, LEP implies a jump in the final direction of the approach to stationary state. The final dynamical direction change at LEP implies LEP could be linked to anomalous phenomena arising in the relaxation process (see Supplementary Table 1).

	\renewcommand{\thetable}{\arabic{table}}
	\begin{table*}[htb]
	\begin{center}
		\caption{ Link between anomalous phenomena in the relaxation dynamics and LEP}
		\label{table:1}
		\begin{tabular}{|c|c|c|c|}
			\hline   \text{Parameter} & $<$\text{LEP} & \text{ = LEP} & $>$\text{LEP} \\
			\hline   Dynamics of $\rho_{in}$ dominate by & $\lambda_1(\in \mathbb{R}), R_1$ & $\lambda_1(R_1)=\lambda_2(R_2),$  & $\lambda_1 = \lambda_2^*,R_1', R_2'$ \\
			\hline   Dynamics of $|\mathrm{sME}\rangle$ dominate by& $\lambda_2(\in \mathbb{R}), R_2'$ & $\lambda_3(R_3)$ & $\lambda_2 = \lambda_1^*, R_2'$\\
			\hline   \text{Exponentially accelerated factor} & $\text{Re}[\lambda_1-\lambda_2]$ & $\text{Re}[\lambda_1-\lambda_3]$ & 0\\
			\hline  \text{Phenomena} & sME & \textit{Super} sME & No sME\\
			\hline
		\end{tabular}
	\end{center}
	\end{table*}
	
	\subsection{Supplementary Note 2. Effective decay rate}
	
	The effective decay channel $\kappa_{1,2}$ from the long life excited state to the ground state in our experiments is created by coupling this long life excited state to another short life excited state, we call it $|p\rangle$, which can quickly decay back to ground state by spontaneous emission process. To describe the full dynamics of this three-level system, we take the decay channel of $|1\rangle \rightarrow |0\rangle$ for example,  the Lindblad master here is
	\begin{equation}
	\frac{d\rho}{dt} = \mathcal L \rho(t):= -i[H,\rho]+ (2J \rho J^{\dagger} - \{J^{\dagger} J, \rho\}),
	\end{equation}
	where $H=\Omega_p/2(|1\rangle \langle p|+|p\rangle \langle 1|)$, $J= \sqrt{\gamma}|0\rangle \langle p|$ is the jump operator describing the spontaneous emission process from $|p\rangle$ to $|0\rangle$ and $\gamma$ is the emission rate.
	
	In order to investigate the physical connotations of the effective decay channel from $|1\rangle$ to $|0\rangle$, we reduce the master equation to the non-zero matrix elements of $\rho$,
	\begin{subequations}\label{subequation1}
	\begin{align}
		\dot{\rho}_{11} =&\ i \frac{\Omega}{2} (\rho_{1p} - \rho_{p1}),\label{suba}\\
		\dot{\rho}_{pp} =&\ -\gamma \rho_{pp} - i \frac{\Omega}{2} (\rho_{1p
		} - \rho_{p1}),\label{subb}\\
		\dot{\rho}_{1p} =&\ -\gamma/2 \rho_{1p} + i\frac{\Omega}{2} (\rho_{11} - \rho_{pp}),\label{subc}\\
		\dot{\rho}_{00} = & \gamma \rho_{pp}.\label{subd}
	\end{align}
	\end{subequations}
	The coherent terms couple the populations $\rho_{11}$ and $\rho_{pp}$ to the coherence $\rho_{1p}$ and $\rho_{p1}$, but have no contribution to the dynamics of coherence $\rho_{1p}+\rho_{p1}$, because $\dot{\rho}_{1p}+\dot{\rho}_{p1} = -\gamma/2(\rho_{1p}+\rho_{p1})$, which is only exponentially damped dynamics with decay rate $\gamma/2$. Meanwhile, as shown in Supplementary Equation.(\ref{subequation1})(a), the damping term, proportional to $\gamma$, does not affect $\rho_{11}$. It contributes to the decay of $\rho_{pp}$ and to the corresponding increase of $\rho_{00}$, and the system eventually relaxes to state $|0\rangle$. Considering there are only three independent variables in Supplementary Equation.(\ref{subequation1}): $\rho_{11}, \rho_{pp}$ and $-i(\rho_{1p}-\rho_{p1})$, we set $x=(\rho_{11},  \rho_{pp}, -i(\rho_{1p}-\rho_{p1}))^T$ and find the dynamics $\dot{x} = A x$, where
	\begin{eqnarray}
	A = \left( \begin{array}{ccc} 0 & 0 & -\Omega_p/2 \\ 0 &-\gamma & \Omega_p/2 \\ \Omega_p/2 & -\Omega_p/2 & -\gamma/2 \end{array}\right).
	\end{eqnarray}\label{supeq7}
	
	Eigenvalues of the $3\times 3$ matrix A in Supplementary Equation (\ref{supeq7}) are $\lambda_{1,2}=-(\gamma \mp \sqrt{\gamma^2-4\Omega_p^2})/2 $ and $\lambda_3=-\gamma/2$. When the coupling driven is strong ($\Omega_p>\gamma/2$), the evolution of $x$ is described by damped oscillation with decay rates $\gamma/2$. When the coupling is weak $\Omega_p<\gamma/2$, all the eigenvalues are real and the dynamics exhibits an irreversible damping with decay rate $-\mathrm{Re}[\lambda_1] = (\gamma - \sqrt{\gamma^2-4\Omega_p^2})/2$. Note that, here our aim is to create an purely exponential decay channel for state $|1\rangle$, so we have to set the Rabi frequency $\Omega_p < \gamma/2$. Specially, in the limit of strong decay or weak coupling, $\gamma\gg 2\Omega$, the effective decay rate of $x$ (of couse also $\rho_{11}$) is determined by
	\begin{eqnarray}
	(\gamma- \sqrt{\gamma^2-4\Omega_p^2})/2 \approx \Omega_p^2/\gamma.
	\end{eqnarray}
	It leads to an effective decay from the state $|1\rangle$ to state $|0\rangle$, with the effective decay rate
	\begin{eqnarray}
	\kappa_1  \approx \Omega_p ^2/\gamma.
	\end{eqnarray}
	The above analysis shows that, this dissipative three-level model can be used to engineer decay processes between state $|1\rangle$ and $|0\rangle$, like the Percell effect in the spin-spring system with relaxation processes \cite{Haroche2010}, just by tuning the Rabi frequency $\Omega_p$. Actually, the same result can be found in \cite{Reiter2012a, Zhang2022} by employing perturbation theory and adiabatic elimination of states $|p\rangle$ for a weakly driven between $|1\rangle \leftrightarrow |p\rangle$.
	
\begin{figure}[htb]
	\includegraphics[scale=0.8]{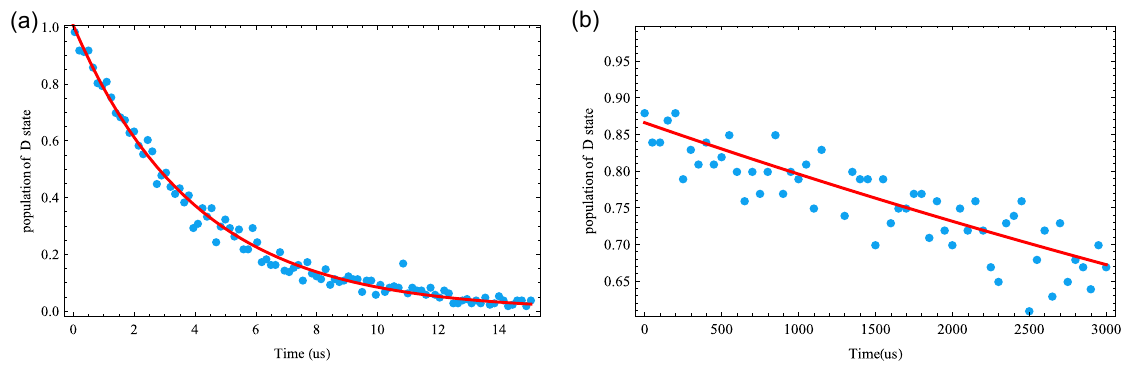}
	\caption{Measurement results for decay rate of states $D_{5/2}(m_j=-5/2)$ and $D_{5/2}(m_j=3/2)$ line. (a) Decay rate $\kappa_1$  for state $D_{5/2}(mj=-5/2)$ with 854 nm laser beam right-handed circularly polarized and  $\kappa_1$ is fitted to be $2\pi\times 39.6 $ kHZ   (b) $\kappa_2$ mesured with same 854 nm laser power and frequency used in (a), $\kappa_2$ is only $2\pi×\times 13.4$ Hz.}
	\label{fig_s5}
	\end{figure}

	Here, we explain how to control and stabilize these decay rates, and further discuss what is the potential impact of this instability on the experimental results.
	Experimentally, the magnetic field of our system is created by using permanent magnets and the left-handed circular polarization of the 854 nm is maintained by using a Glan-Taylor prism and a quarter waveplate, hence we consider these parameters are stable during our experiments. Then we set the frequency of the 854 nm  to be resonant with line $P_{3/2}(m_j=-3/2) \leftrightarrow D_{5/2}(m_j=-5/2)$ so that other lines are off-resonant and the decay rate can be suppressed to some extent because of the different Lande g factors for $D_{5/2}$ state and $P_{3/2}$ state. Finally, we control the decay rates of the three dipole transition lines to by setting the polarization of the 854 nm.  For the line  $P3/2(m_j=-3/2) \leftrightarrow D5/2(m_j=-5/2)$, the beam of 854 nm should be right-handed circularly polarized, while the 854 nm laser beam for lines $P_{3/2}(m_j=1/2) \leftrightarrow D_{5/2}(m_j=3/2)$ and $P_{3/2}(m_j=3/2) \leftrightarrow D_{5/2}(m_j=3/2)$ are left-handed circularly polarized and linearly polarized, respectively. Therefore, we make the 854 nm light polarization to be right-handed circularly polarized, and the wavevector of the 854 nm to be almost perfectly along with the magnetic field after careful alignment. In this case, we suppressed $\kappa_2$ significantly. The example measurements for $\kappa_1$ and $\kappa_2$  are shown in Fig. \ref{fig_s5} (a) and Fig. \ref{fig_s5} (b) below.

	\bibliography{bibfile}
	
\end{document}